\documentstyle[11pt,aaspp4,fleqn]{article}

\begin{document}


\title{POLARIMETRIC PROPERTIES OF THE CRAB PULSAR\\ BETWEEN 1.4 AND 8.4 GHz}
\author{David A.~Moffett\altaffilmark{1} and Timothy H.~Hankins}
\affil{Physics Department, New Mexico Institute of Mining and
Technology\\ Socorro, NM 87801}

\altaffiltext{1}{
Current Address: School of Mathematics and Physics, University of Tasmania, 
GPO Box 252-21, Hobart 7001, Australia}

\begin{abstract}
New polarimetric observations of the Crab pulsar at frequencies
between 1.4 and 8.4\,GHz are presented. Additional pulse components
discovered in earlier observations (\cite{mh96}) are found to have
high levels of linear polarization, even at 8.4\,GHz.  No abrupt
sweeps in position angle are found within pulse components; however,
the position angle and rotational phase of the interpulse do change
dramatically between 1.4 and 4.9\,GHz.  The multi-frequency profile
morphology and polarization properties indicate a non-standard origin of
the emission.  Several emission geometries are discussed, but the one
favored locates sites of emission both near the pulsar surface and in the
outer magnetosphere.
\end{abstract}


\section{Introduction}

Past studies of the polarization properties of the Crab Nebula pulsar
have been limited to low-frequency radio and visible wavelengths.  The
pulsar's steep radio spectrum, and interference from the radio-bright
Nebula make observations above 1\,GHz difficult with single dish
antennas.  Thus, interpretations of the pulsar's emission geometry 
have only been made from the properties of its polarized profiles at
visible wavelengths.

Single dish average profile measurements of the radio polarization are
available at frequencies between 110 and 1664\,MHz (Manchester,
Huguenin, \& Taylor 1972\nocite{mht72}; Manchester
1971a\nocite{man71a}).  Prior to the work described in Moffett \&
Hankins (1996, hereafter Paper I), only three components of the
pulsar's average profile were known; a steep-spectrum precursor, which
is approximately 100\% linearly polarized (\cite{man71a}), plus a main
pulse (MP) and interpulse (IP) which are roughly 15 to 25\% linearly
polarized.  The position angle (PA) remains constant between all three
components, with very little change of position angle across them.
The only other major radio polarization observations that have been
published are measurements of the time-variable rotation measure used
to probe the magnetic fields of the Crab Nebula's filaments
(\cite{rci+88}).  No high radio-frequency ($\nu > 1.7$\,GHz)
polarization information for the Crab pulsar has been available.

The visible wavelength (\cite{sjd+88}) and newly acquired ultraviolet
(\cite{sdb+96}) polarization profiles show similar linear polarized
fractions as the radio, but unlike the radio profiles, they show large
PA variations.  At the peak positions of the MP and IP, the fraction
of visible linear polarization is about the same as in the radio
regime, $\approx 14$ to 17\%. In the region after the IP in phase, the
percentage polarization rises to $\approx 47 \pm 10$\%, and the
position angle rises above the IP angle and remains nearly constant
across the total intensity minimum.  Narayan \& Vivekanand
(1982)\nocite{nv82} found that the visible wavelength polarization PA
sweeps of the MP and IP suggest that emission comes from two opposite
poles of a pulsar whose magnetic axis is nearly orthogonal to the
rotational axis.  But arguments from $\gamma$-ray emission theory have
surfaced recently (Manchester 1995\nocite{man95}; Romani \&
Yadigaroglu 1995\nocite{ry95}) that question this type of geometry by
claiming that emission arises from a wide cone in the outer
magnetosphere.

So far, the study of radio polarization has not improved our knowledge
of the emission and field geometry.  The lack of position angle
variation in low frequency radio profiles is difficult to explain in
terms of the simple rotating vector model (\cite{rc69}).  Following
the serendipidous discovery of additional components in the Crab
pulsar's profile in Paper I, a program of polarization observations
was scheduled to study the high radio-frequency polarization
characteristics of these new components, perhaps improving the
interpretation of the polarization and emission location for the Crab
pulsar's radio components.

\section{Observations}

Observations were conducted during several sessions from October 1995
to October 1996 at the Very Large Array (VLA) of NRAO.  
Between February 22 and April 18, 1996, the data acquisition system
was modified to double the number of filterbank channels that are
recorded.  Using the phased VLA, the coherent sum of undetected
right-hand and left-hand circular polarization ($R$ and $L$) from all
antennas is mixed to 150\,MHz and then split into 14 independent
frequency channels by a MkIII VLBI filter bank.  The filter bank
output is sent to the VLA's High Time Resolution Processor (HTRP),
which consists of a set of 14 multiplying polarimeters.  Channels of
detected and smoothed $LL$, $RR$, $RL\cos{\theta}$, and $RL\sin{\theta}$, where
$\theta$ is the phase offset between $R$ and $L$, are continuously
sampled by 12-bit, analog-to-digital converters in a PC and recorded
on disk at a time resolution of 256 $\mu$s.  The detector time
constants are set to optimize sampling of the dispersed time series
across the channel bandwidths.

The observations were scheduled so that short scans (typically 30 to
40 minutes apart) of an unresolved calibrator point source were made
between pulsar scans to keep the VLA phased, and to record on- and
off-source data for flux and polarization calibration of the pulsar
data.  Within the duration of an observing session, anywhere from five
to seven sets of measured Stokes parameter fluxes were recorded from
the phase and flux calibrator, 3C138, with enough parallactic angle
coverage for instrumental polarization calibration.  The flux density
and position angle of 3C138 are regularly monitored by the University
of Michigan Radio Astronomy Observatory.  The position angle of this
source remains the same over our frequency coverage, with a value of
$\psi = -12^\circ$.

The pulsar data were folded off-line at the pulsar's topocentric
period using a timing model initially provided by Nice (1995).
Consequent observations of the Crab pulsar using the
Princeton/Dartmouth Mark III Pulsar Timing System (\cite{skn+92})
provided time-of-arrival (TOA) information reduced using the program
TEMPO (\cite{tw89}), which yielded new timing solutions for folding at
later epochs.  Individual channel data were folded into two-minute
average profiles of all four detected polarizations prior to
calibration and dedispersion.

The gain amplitudes, relating the received voltage in the data
acquisition system to flux density in Janskys for the $LL$ and $RR$
detector signals were determined by observation of the phase
calibrator source and blank sky.  The gain amplitude of the cross
polarizations, $RL\cos{\theta}$ and $RL\sin{\theta}$, were found
directly from solutions for the circular polarization gains $G_{\rm
  L}$ and $G_{\rm R}$ from $LL$ and $RR$, by using $G_{\rm RL} =
\sqrt{G_{\rm R} \cdot G_{\rm L}}$.  For data collected from the lower
side-band channels of the MkIII VLBI videoconverters, the sign of the
measured Stokes U was inverted, thus removing a known $180^\circ$
phase shift caused by the image-rejecting mixers within the
videoconverters.

Polarization calibration was completed following procedures similar to
those used by McKinnon (1992)\nocite{mck92}.  In his paper, the
polarization characteristics of the phased VLA approximate those of a
single dish antenna with circular polarization feeds.  An ideal
antenna with orthogonal circular polarization receivers has no
cross-coupling.  However, imperfections in the reflectors and
receiving systems of antennas tend to change the received radiation
from purely linear polarized sources into elliptical polarization
(\cite{ck69}).

McKinnon's method involves measuring the time-dependent Stokes
parameters of a polarization calibrator source with respect to the
changing parallactic angle, and solving for time-dependent and
independent instrumental corrections.  McKinnon used the polarization
from a point in a pulsar's profile to perform a self-calibration, but 
he could not determine the absolute position angle.  We could have 
used the Crab pulsar at 1.4\,GHz to perform such a self-calibration, 
but at higher frequencies we were limited by low signal to noise ratios.  
Instead we used a phase calibrator of known polarization 
characteristics to solve for the instrumental corrections {\em and} 
absolute position angle at 1.4, 4.9, and 8.4\,GHz.

\section{Results}

Our results at 1.4\,GHz (Figure \ref{lband_fig}) are similar to
published observations at 1.664\,GHz (Manchester 1971a\nocite{man71a};
Manchester 1971b\nocite{man71b}), but with the addition of three more
components (see Paper I, Figure 2).  These three are labeled LFC for
``low frequency component'', as it mainly appears at $\nu < 2$\,GHz,
and HFC1 and HFC2 for ``high frequency component 1 and 2'', as they
appear only at $\nu\geq 1.4$\,GHz.  The MP and IP are both linearly
polarized, 25\% and 15\% respectively, and their relative PAs are
nearly the same.  The LFC component is more than 40\% linearly
polarized, and it has a PA offset $\approx 30^\circ$ from the MP,
which sweeps down toward the MP.  There is also low-level emission
after the IP, coincident in phase with components HFC1 and HFC2 found
in the total intensity profiles at 4.9\,GHz (Paper I). These components
are $> 50\%$ polarized, and their position angles appear to be
relatively the same, offset from the MP by $60^\circ$.

The circular polarization undergoes a sense reversal centered on the
MP with an amplitude 1 -- 2\% of the MP peak.  We can exclude this as
a cross-coupling signature, even though our uncertainties for fitting
the instrumental parameters were several times higher for individual
frequency channels.  The linear polarization is not strong, nor does
it sweep rapidly across the pulse at our time resolution.  So a
coupling of linear to circular power should not produce
sense-reversing circular polarization.  We can make no comparisons as
no previous circular polarization observations of average profiles
have been published.  The similarity of circular polarization
signatures on separate observation dates, and in individual channels
improves our confidence that these signatures are real.

\placefigure{lband_fig}

The profiles at 4.9\,GHz (Figure \ref{cband_fig}) are
new results.  We have successfully confirmed the detection of the HFC
components found in Paper I.  The pulsar was visible
for only three out of seven observing sessions; the profile was formed
from about 3.7 hours of data.  We attribute the non-detections to
heavy scintillation, which affected observations at 
4.9\,GHz and higher frequencies.  The IP, HFC1, and HFC2 are highly
polarized, 50\% to 100\%, while the MP seems to have the same
polarized fraction as at 1.4\,GHz.  HFC1 and HFC2 share a
common range of PA, but it sweeps through them with different slopes.
The most important feature to note is the IP.  Its relative flux
density has increased with respect to the MP and it has shifted
earlier in phase by $\approx 10^\circ$ (see Paper I, Figure 2).

\placefigure{cband_fig}

At 8.4\,GHz (Figure \ref{xband_fig}), the profiles show some
polarization, and also confirm the profile morphology found in earlier
total intensity observations.  The pulsar was visible for only one out
of three observing sessions for a total of 2.3 hours, which we again
attribute to scintillation.  The IP seen in the profile is
substantially wider than at lower frequency and the fractional
polarization of the IP and HFCs is reduced, due in part to an
incomplete instrumental correction of the position angles.  As in
Paper I, no evidence is found of the MP, whose predicted flux from a
spectral index of $\alpha_{\rm MP} = -3$ (Section \ref{spix_sec}) is
below the noise level of the profile recorded at this frequency.

\placefigure{xband_fig}

Observations made on April 18, 19 and 20, 1996, were conducted at
several frequencies within the 1.4\,GHz receiver band, and evidence
for Faraday rotation was found between the separate frequencies.  A
rotation measure, RM$ = -46.9$\,rad\,m$^{-2}$, was found after
comparing the position angles of the major components, and its effects
have been removed from all profiles reported here.  Past measurements
show the RM near $-43.0$\,rad\,m$^{-2}$ (\cite{rci+88}), but
it is known to be variable on time scales of months, as the line of
sight to the pulsar passes through the Crab Nebula's filaments.
After removing Faraday rotation effects, the position angles of the
MP, HFC1 and HFC2 are found to align at 1.4 and 4.9\,GHz, and the PAs
of the IP, HFC1 and HFC2 align at 4.9 and 8.4\,GHz.  But, the IP is
found to have a position angle difference between 1.4 and 4.9\,GHz of
$90^\circ$ (see Figure \ref{angle3freq_fig}).  So the IP has a
discontinuous change of positional phase, flux, and polarization
between 1.4 and 4.9\,GHz.  It obviously cannot be the same component
at both frequencies.

\placefigure{angle3freq_fig}

\section{Analysis}

The unique and confusing discoveries described in the previous section
are the first successful, fully polarimetric observations of the Crab
pulsar above the 1.4-GHz band.  In the following sections, this pulsar's 
emission geometry is explored by comparing properties of the its 
polarization profile with known properties of other pulsars, and 
possible emission geometry models.

\subsection{Multiple Components}

The components of the Crab pulsar appear in six distinct positions in
rotational phase at all observed radio frequencies.  The distribution 
of components is difficult to
explain in a low-altitude, dipolar or hollow-cone emission models
(\cite{ran83a}; \cite{lm88}), mainly because of their number and wide
separation.  Up to 5 components have been seen from ``normal'' pulsars 
(PSR B1237+25 and B1857-26).  The separation of profile
components is usually restricted to a small range of pulse phase ($<
30^\circ$), corresponding to a cone of emission above one pole of the
star.  However, a few interpulsars exist, whose components can be
attributed to emission from the observer's line of sight passing above
both poles (orthogonal rotator), or from one pole (aligned rotator).

The phase separation between the MP and IP, $\Delta\phi_{\rm MP-IP}
\approx 140^\circ$, is too low to argue for a line of site crossing of
both poles.  When compared to the high energy emission (infrared to
$\gamma$-ray), the morphology of the MP and IP implies that they arise
from a wide conal beam, high above a single pole (\cite{man95}).  With
the wide beam picture in mind, one apparent symmetry in the
distribution of the Crab's components can be seen if we draw a line
through the midpoint between the MP and IP, and the midpoint between
HFC1 and HFC2 (see Figure 3 of Paper I).  The midpoints between the
component pairs are separated by $\approx 170^\circ$ at 4.9\,GHz.  So
the Crab's components could arise from conal emission regions above
both poles, one wider than the other.  In fact, the HFCs do show
promise as a conal pair.  Rankin's (1993)\nocite{ran93a} empirical
relations for inner and outer conal width (assuming the Crab to be an
orthogonal rotator, $\alpha = 90^\circ$) yield:
\begin{displaymath}
\begin{array}{cclcl}
\rho_{\rm inner} & = & 2 \times 4.33^{\circ}\,P^{-1/2} = 47.3^{\circ}\\
\rho_{\rm outer} & = & 2 \times 5.75^{\circ}\,P^{-1/2} = 62.8^{\circ}
\end{array}
\end{displaymath}
The phase separation of the HFCs, $\Delta\phi_{\rm HFC1-HFC2} \approx
56^\circ$, is within Rankin's predicted values for inner and outer
conal widths for a pulsar of the Crab's period.  It is possible 
that the HFCs are generated at low altitudes, and the MP and IP 
are generated at higher altitudes where the emission beam is much wider.
However, we note that interpreting the frequency-dependent properties 
of the IP and the HFCs with this geometric model is quite difficult.

Another set of components, the LFC, precursor, and main pulse, form
what may be a cone/core triplet.  The LFC to MP separation is $\approx
45^\circ$, nearly what one expects for the inner conal width, and the
precursor behaves much like a core component, with its high polarization 
and steep spectrum.  But why the MP is so much brighter than the LFC 
requires explanation.

It is interesting to note that one pulsar, B1055-52, has a similar
distribution of components (precursor, main pulse, and a strong
interpulse located $155^\circ$ away) at low frequency (\cite{mha+76}).
And like the Crab, it also has pulsed high energy emission X-rays
(\cite{of93}), pulsed $\gamma$-rays (\cite{fbb+93}), and has been
recently detected as continuum source at visible wavelengths
(\cite{mcb97}).

\subsection{Radius to Frequency Mapping? \label{freq_dep_sec}}

Using the main pulse as the fidicial point of the Crab's profile
(Paper I), we found that the separations from MP to IP, and from MP to
the HFCs are frequency-dependent.  (see Figure \ref{phase_sep_fig}).
From 1.4 to 4.7\,GHz, the IP jumps $\approx 10^\circ$ earlier in
phase, while the HFCs appear to make a smooth linear transit in phase
between 1.4 and 8.4\,GHz.  This property is reminiscent of the smooth
phase shift of conal components in radius-to-frequency mapping (Cordes
1978\nocite{cor78}; Rankin 1983b\nocite{ran83b}; Thorsett
1991\nocite{tho91}).  The phase separation of conal components usually
can be best fit by a power law function, $\Delta\phi \propto
\nu^{\eta}$, where $\-1.1 \leq \eta \leq 0.0$.  The phase separations
from the MP to both HFC1 and HFC2 are best fit with $\eta = 1$ (fit
parameters found in Figure \ref{phase_sep_fig}).  The HFCs are both
moving toward later rotational phase with increasing frequency, unlike
conal components of other pulsars, whose phase separation decreases to
a common fiducial point.  Curiously, the HFCs would merge at a common
point at the MP phase, if their phases are extrapolated to a frequency
above 60\,GHz.

\placefigure{phase_sep_fig}

\subsection{Spectral Index\label{spix_sec}}

The amplitude calibration method for these observations was based on
gains transferred from a standard extragalactic continuum calibration
source, whereas the flux densities in the profiles presented in Paper
I were estimated using known radiometer characteristics.  We used the
integrated flux density under the major components, and computed
spectral indices for the MP and IP; $\alpha_{\rm MP} = -3.0$ for the
MP, and $\alpha_{\rm IP} = -4.1$ for the IP at $\nu \leq 1.4$\,GHz.
Independent of the uncertainty of the flux density measurements, the
relative spectral index differences between components were determined
simply through ratios of their integrated flux densities using the
following relation:
\begin{displaymath} 
{S_{\rm C1}(\nu_1)/S_{\rm C1}(\nu_2) \over S_{\rm C2}(\nu_1)/S_{\rm C2}(\nu_2)}
= {\nu_1 \over \nu_2}^{(\alpha_{\rm C1} - \alpha_{\rm C2})} 
\end{displaymath}
where the fluxes, $S_{\nu}$, and spectral indices, $\alpha$,
correspond to the components C1 and C2.  The spectral index difference
of components using these ratios yields a spectral index, $\alpha_{\rm PC} 
\simeq -5.0$, for the precursor.  The spectral indices we have
found for the three major components of the Crab pulsar profile agree
with previous measurements by Manchester (1971a)\nocite{man71a}.

In Figure \ref{flux_vs_freq_fig}, we plot the flux density spectrum of
the MP and IP, and the two HFC components.  Below 1.4\,GHz, the IP
follows a power-law spectral index of approximately -4, but above
1.4\,GHz, the plot shows that the IP has a flat spectral index, as do
the HFC components, though no power-law can be determined from the
plot.  Such a turn-up or flattening of pulsar spectra has been
observed by Kramer {\it et al.} (1996) \nocite{kxj+96} in two other
pulsars.  They have suggested that a transition from coherent to
incoherent emission would cause changes in the expected flux density.
Sampling pulsar radiation at very high frequencies gives limits to the
bandwidth of the coherent emission mechanism.  A simple extrapolation
of the Crab pulsar spectrum from radio to infrared wavelengths (Fig.
4-2, Manchester \& Taylor 1977\nocite{mt77}) implies that the flux
must rise and the emission mechanism must change.  So the change in 
spectral index lends support to our hypothesis that the low frequency 
IP and high frequency IP are two different components.

We should note, that when compared to other pulsars, the
spectral indices of the Crab pulsar's MP and IP are much steeper
than the components of other pulsars ($-1.5 < \alpha < -3$).  
The Crab's mean spectral index, $\alpha_{\rm crab} = -3.1$, is also 
greater than the average spectral index, $\alpha = -1.5$, of most 
detected pulsars (\cite{lyl+95}).

\placefigure{flux_vs_freq_fig}

\subsection{Polarization Properties}

The polarization position angle of the Crab changes across the full
period, though not significantly within components.  There are no
sudden well-defined PA sweeps (`S' shaped sweeps) within components,
as seen at optical wavelengths.  However, we should note the radio
components are much narrower, and some polarization information is smeared
by dispersion and scattering.  The lack of PA variation between close
components implies that the observer's line of sight trajectory does
not fall close to the magnetic poles, where the position angle of
field lines varies quickly.

The fraction of linear to total intensity of the MP and HFCs is nearly
constant from 1.4 to 4.9\,GHz.  But the IP becomes substantially more
polarized (from 20\% to 100\%) between the two frequencies, as well as
undergoing a $90^\circ$ PA shift.  The spectral change in phase and PA
could be due to a mechanism (birefringence) affecting the propagation
of the two orthogonal modes (ordinary or O-mode, and extraordinary or
X-mode) of linear polarized radiation within the pulsar's
magnetosphere (\cite{ba86}).  The ordinary mode waves are forced to
travel along magnetic field lines, while the extraordinary mode waves
are unaffected.  A sudden change in plasma conditions could cause one
of the modes to be beamed out of the line of sight.  However, this
process is sensitive to frequency, and any transitions we see should
be continuous.  The change of the phase and PA of the IP between 1.4
and 4.9\,GHz is rather abrupt, but this does not rule out
birefringence effects, since we have not yet seen the IP at an
intermediate frequency (\cite{mh96}).

In general, the polarized fraction of other pulsars {\em decreases}
with frequency, and the position angle gradient is independent of
frequency (\cite{xkj+96}).  It is generally believed that pulsar {\em
depolarization} toward higher frequencies is due to the instantaneous
superposition of emitted orthogonal or quasi-orthogonal polarization
modes (\cite{scr+84a}).  Our results seems to indicate that one
polarization mode dominates the emission from the IP and HFCs for 
$\nu > 1.4$\,GHz.

Following the standard rotating vector model (RVM), the polarization
position angle traces the projected magnetic field of the pulsar if
emission occurs along the open field lines. The position angle of the RVM is given by Manchester \& Taylor (1977)\nocite{mt77} as
\begin{equation}
\psi(\phi) = \psi_0 + {\rm tan}^{-1} 
\left[ {\sin{\alpha} \sin{(\phi - \phi_0)} \over 
\sin{\zeta}\cos{\alpha} - 
\cos{\zeta}\sin{\alpha}\cos{(\phi - \phi_0)}} \right] \, ,
\label{RVM_eqn}
\end{equation}
where $\psi_0$ is a position angle offset, $\phi_0$ is the pulse phase
at which position angle variation is most rapid, $\alpha$ is the
inclination angle from the rotation axis to the magnetic axis, and
$\zeta$ is the angle between the rotation axis and the observer's
line-of-sight.  The observer's impact angle with the magnetic axis is
just the difference $\beta = \zeta - \alpha$. It can also be
determined from the maximum slope of position angle with phase by using
\begin{equation}
\left[{d\psi \over d\phi}\right]_{\rm max} = {\sin{\alpha} \over \sin{\beta}}
\label{impact_eqn}
\end{equation}
This simple geometric construct is only useful if emission is located
close to the polar cap, since the line of sight angle $\zeta$ in the
model passes through the center of the star (which is not the case in
reality).

We have made rudimentary fits of Eq(\ref{RVM_eqn}) to our polarization
profiles in an attempt to match polarization signatures to the low
altitude dipole model.  In Figure \ref{pafit_fig}, we plot the
position angles of the Crab's major components at 1.4, 4.9 and
8.4\,GHz with the best fit to the RVM overplotted. The RVM does not
fit well for the case where the PA of the IP at all frequencies is
left at its 1.4-GHz value, so we have shifted the IP position angle at
1.4\,GHz by $90^\circ$ to match the that at higher frequency.  From
the fit, the angle found between the rotation and magnetic axes is
$\alpha = 56.0^\circ$, with one pole projected near the IP, and the
other near the LFC.  With $\alpha$ fixed in Eq(\ref{impact_eqn}), the
slope of the position angle with phase yields an impact angle, $\beta
\simeq 51^\circ$, for the IP.  The LFC has a smaller impact angle,
$\beta \simeq -30^\circ$, and the maximum slope of the fitted curve
occurs just ahead of the LFC.  At this location, $\beta \simeq
-18^\circ$.  The impact angles for the components near the interpulse
are much larger than those found from other pulsars (Rankin
1993b\nocite{ran93b}; Lyne \& Manchester 1988\nocite{lm88}).  The
fitted impact angles are much larger than the polar cap width expected
for the Crab pulsar, given by Goldreich \& Julian (1969)\nocite{gj69}
is $\rho_{\rm PC} = \left( \frac{2\pi r }{c P} \right)^{-1/2} =
4.56^\circ$, where $\rho_{\rm PC}$ is the width of the polar cap, $r$
is the height above the surface, $c$ is the speed of light, and $P$ is
the pulsar's rotational period.  So our solution to the RVM fit
appears to find emission both close to (LFC) and well away from 
(IP, HFC1 and HFC2) expected low altitude dipole fields above the polar cap.

\placefigure{pafit_fig}

Our radio wavelength RVM fit does not agree with values found from
fitting the visible wavelength PA sweeps (\cite{nv82}): $\alpha =
86^\circ$, $\beta_{\rm MP} = 9.6^\circ$, and $\beta_{\rm IP} =
-18^\circ$.  The large inclination angle and the small observer impact
angles of this fit imply that both poles sweep by the observer.  However, the
visible wavelength fits were obtained by only fitting for the maximum
sweep through each component individually, not by fitting the position
angle over the whole pulsar period.

A simple comparison of the visible polarization profiles (\cite{sjd+88}) and
the high frequency radio profiles show a few similarities.  First, the
PA of the MP and IP at 1.4\,GHz matches the visible PA at the phase of
the radio components.  And though PAs of components at high radio frequency do
not match the visible, the position angles and polarized fraction of 
both the visible and radio profiles increase in the region occupied by 
HFC1 and HFC2.

\section{Emission Geometry}

It is difficult to interpret the emission geometry from the profile
morphology and polarization measurements we have acquired.  There are
six sites in rotational phase where pulsed emission occurs, and some
evidence of radius to frequency mapping.  The HFC components appear to
be separated by a width comparable to the conal width expected of a
pulsar of this period (Rankin 1993), as do the LFC and MP pair.  
The precursor may even be a core-type component between the LFC and MP.
However, the sweep of position angle through these components is shallow,
suggesting that radiation comes from far outside a low-altitude emission cone.

So far, our interpretation has followed a simple emission geometry
proposed by Smith (1986)\nocite{smi86}, which places the location of
emission at both low and high altitudes.  The MP and IP are generated
in the outer magnetosphere, near the light cylinder, where the dipolar
fields are swept back, and the rotational phase of components and
their position angles is not the same as above the polar cap.
Although no clear evidence of field sweep back has been found for
pulsars, if the emission does originate at high altitudes, the
swept-back dipole fields of the pulsar would allow the MP and IP to be
formed from either the two sides of the same dipole cone above one
pole, or from just the leading edges of dipolar fields above both
poles of an orthogonal rotator (\cite{sjd+88}).  The precursor and LFC
are then generated close to the surface of the star above one pole.
However, this simplistic model does little to interpret the HFC
components, how the IP's properties change, or the nature of the
polarization position angle.

Another model, proposed by Romani and Yadigaroglu (1995)\nocite{ry95},
ties $\gamma$-ray emission of several pulsars to particle production
in an outer magnetospheric gap.  Through Monte Carlo simulations of
particles in the gap, they have generated $\gamma$-ray profiles
similar to the Crab and Vela pulsars, and have successfully generated
a polarization position angle profile similar to that of the optical
polarization of the Crab, by projecting the magnetic fields (or
polarization of high energy photons) in the outer gap.  Using this
model, it is even possible to find a less powerful outer gap surface
that could drive particle acceleration at rotational phases where HFC1
and HFC2 reside (\cite{rom96}).  The processes by which radio
radiation is generated in the outer magnetosphere are still unknown,
though they must be similar to normal pulsar radio production to yield
comparable radio power and spectra.  Romani and Yadigaroglu
(1995)\nocite{ry95} also claim that one should see low altitude
emission alongside the outer magnetospheric emission if the
orientation of the pulsar allows it.  This is true for the 
Crab pulsar's precursor as well as the Vela pulsar's single radio 
component, which is offset in phase from its X-ray emission.  
%

One last piece of information that may aid in efforts to interpret the
polarization, is evidence for the Crab pulsar's orientation on the
sky.  Using optical images from HST, \cite{hss+95} link certain
features found at visible wavelengths with structures found in ROSAT
X-ray images.  The wisps, arcs, and jet-like features, which probably came
from interactions of the Nebula with a pulsar wind, show a cylindrical
symmetry, implying that the spin axis of the pulsar is at an angle of
$110^\circ$ east of north, projected $\approx 30^\circ$ out of the
plane of the sky to the southeast.  If the geometry proposed by Hester
{\it et al.} is tied to the true spin axis of the pulsar, then the
angle of the spin axis to the observer is $\alpha = 90^\circ -
30^\circ = 60^\circ$, very close to our fitted value for the observer
impact angle to the spin axis determined through RVM fits.

\section{Conclusion}

We have presented new polarimetric observations of the Crab pulsar at
frequencies between 1.4 and 8.4\,GHz which are difficult to interpret
under the classical polar cap model.  There are more than the typical
number of components seen in other pulsars, and they arise from all
over the pulsar's rotational phase. The new pulse components (LFC,
HFC1 and HFC2) found in Paper I all have high linear polarization.  We
re-confirmed the phase shift and spectral change of the IP between 1.4
and 4.9\,GHz, and found that the component also undergoes a $90^\circ$
relative position angle shift with respect to the other components!  A
good fit is made of the low altitude, rotating-vector model to the
polarization position angle at high frequencies, but the line of sight
impact angles to the magnetic axis are very large, implying that
emission is arising from angles beyond the width of the low altitude
polar cap region.  It appears that the MP and IP do not arise from low
altitude dipole emission.  However, the LFC and HFC components show
some properties inherent to conal emission (just as the precursor
exhibits core-type emission).  The Crab profile appears to be
associated with a mixture of low and high altitude emission, with the
IP being the greatest mystery and exception to the rule.

\acknowledgments

The authors wish to thank Phil Dooley and the LO/IF group at
NRAO-Socorro for their maintenance of the HTRP system.  DM
acknowledges support from a NRAO pre-doctoral fellowship, and from NSF
grants AST 93-15285 and AST 96-18408.  The National Radio Astronomy
Observatory is a facility of the National Science Foundation operated
under cooperative agreement by Associated Universities, Inc.  This
research has made use of data from the University of Michigan Radio
Astronomy Observatory which is supported by the National Science
Foundation and by funds from the University of Michigan.  We also
acknowledge the use of NASA's Astrophysics Data System Abstract
Service (ADS), and the SIMBAD database, operated at CDS, Strasbourg,
France.



\epsscale{0.75}
\plotone{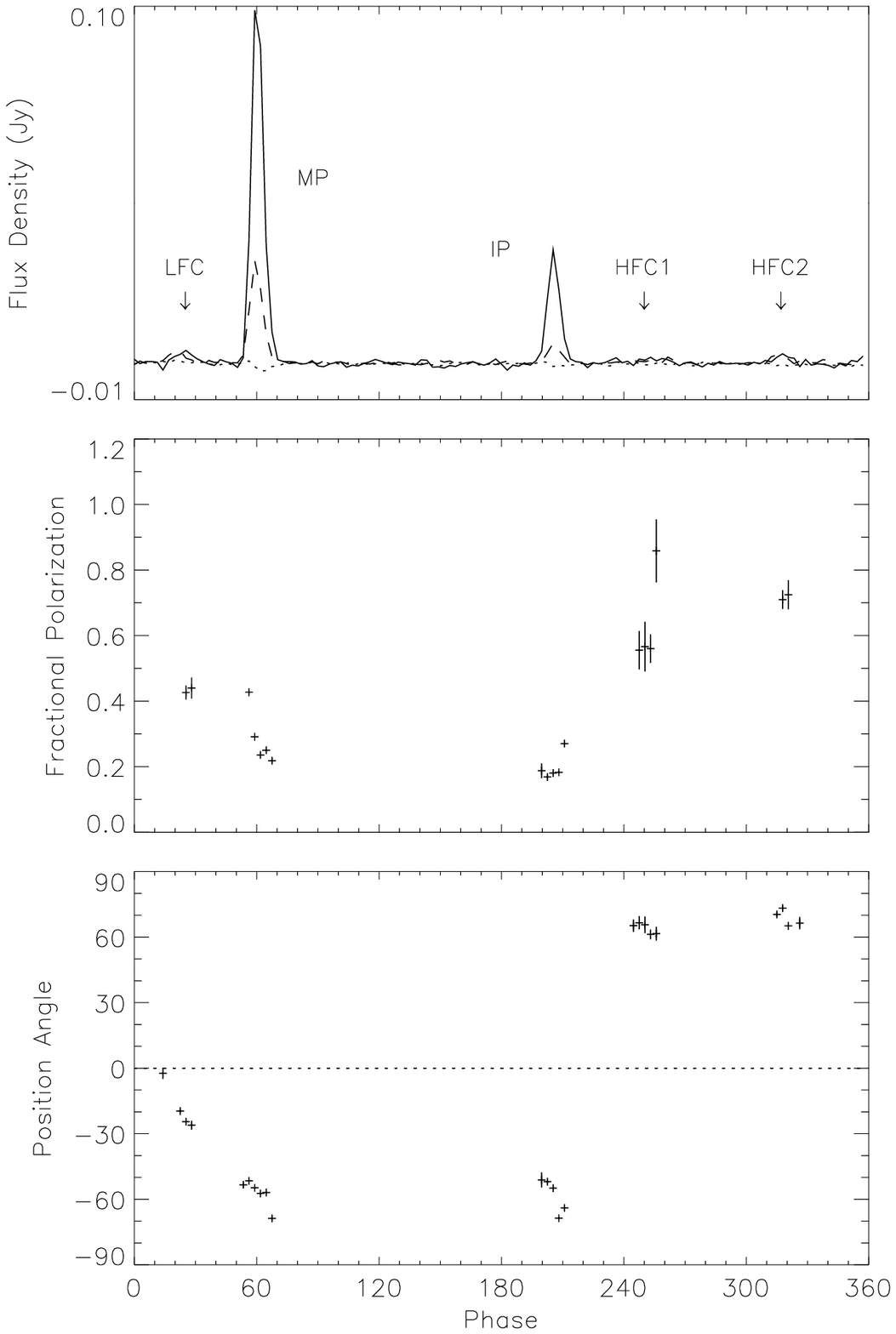}
\figcaption[f1.ps]{Polarization profile of the Crab at 1.424\,GHz.  
Top: total intensity (solid), linear polarization (dashed), and circular 
polarization (dotted).  
Middle: fractional polarization.  
Bottom: position angle of
the electric vector after Faraday rotation correction. Fractional
polarization and position angle plotted only for points above
$5\sigma$ of the off-pulse noise.\label{lband_fig}}

\newpage
\epsscale{0.75}
\plotone{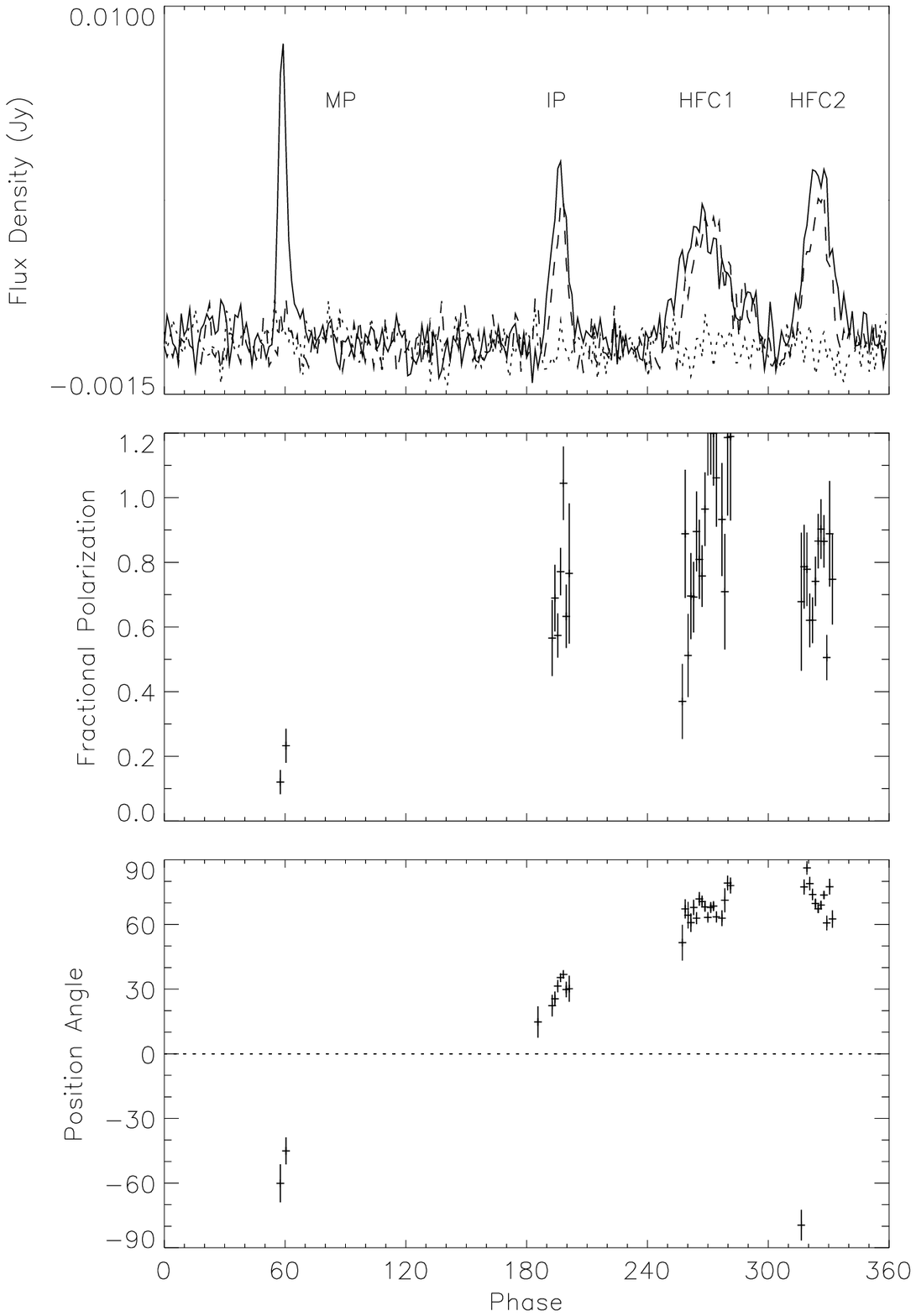}
\figcaption[f2.ps]{Polarization profile of the Crab at 4.885\,GHz.
Top: total intensity
(solid), linear polarization (dashed), and circular polarization (dotted).  
Middle: fractional polarization.  
Bottom: position angle of
the electric vector after Faraday rotation correction.  
Fractional polarization and
position angle are plotted only for points above $3\sigma$ of the
off-pulse noise.\label{cband_fig}}

\newpage
\epsscale{0.75}
\plotone{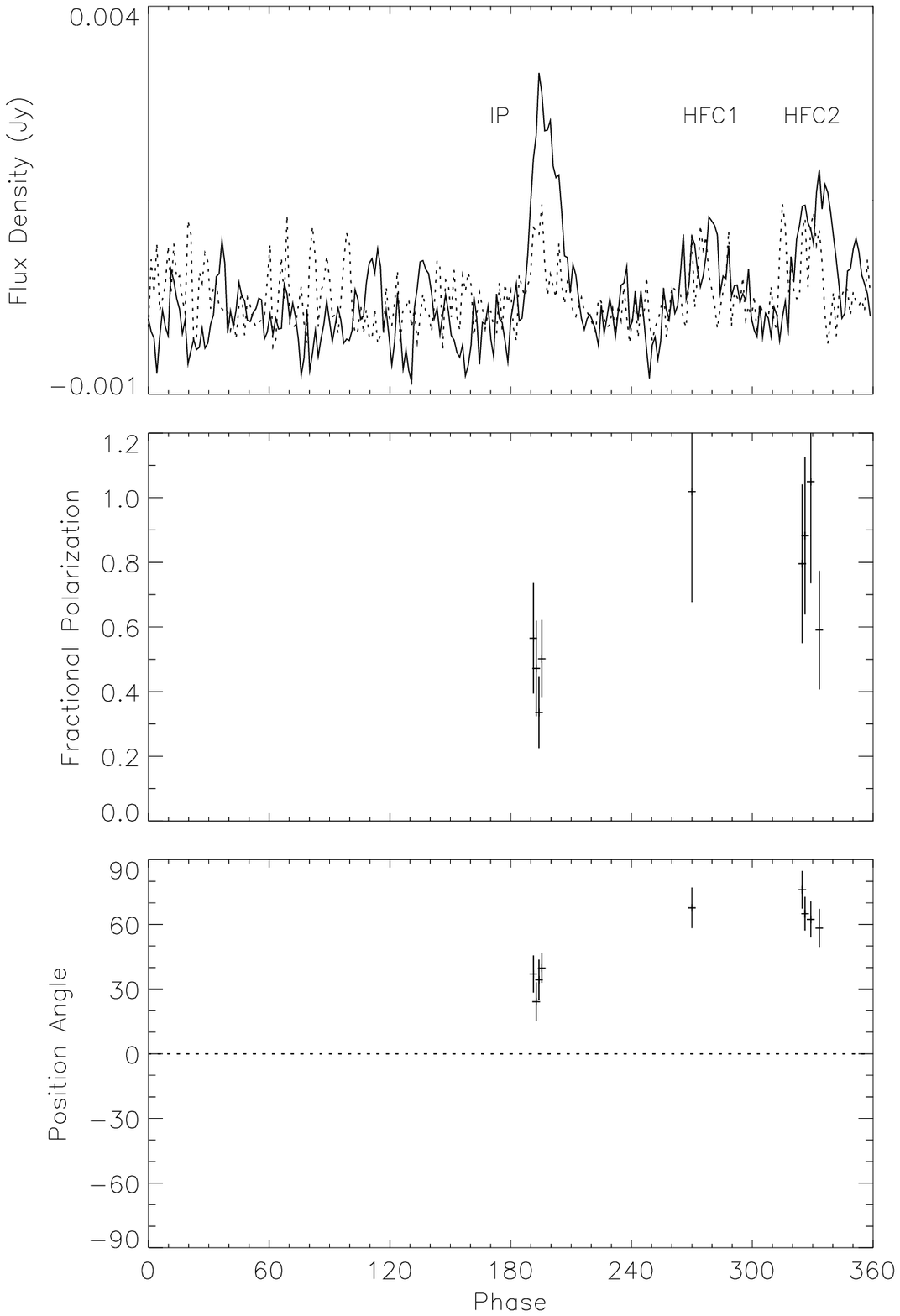}
\figcaption[f3.ps]{Polarization profile of the Crab at 8.435\,GHz.
Top: total intensity (solid) and linear polarization (dotted).
Middle: fractional polarization. 
Bottom: position angle of the electric vector after Faraday rotation 
correction.
Polarization and position angle are plotted only for points above 
$3\sigma$ of the off-pulse noise.\label{xband_fig}}

\newpage
\epsscale{1.0}
\plotone{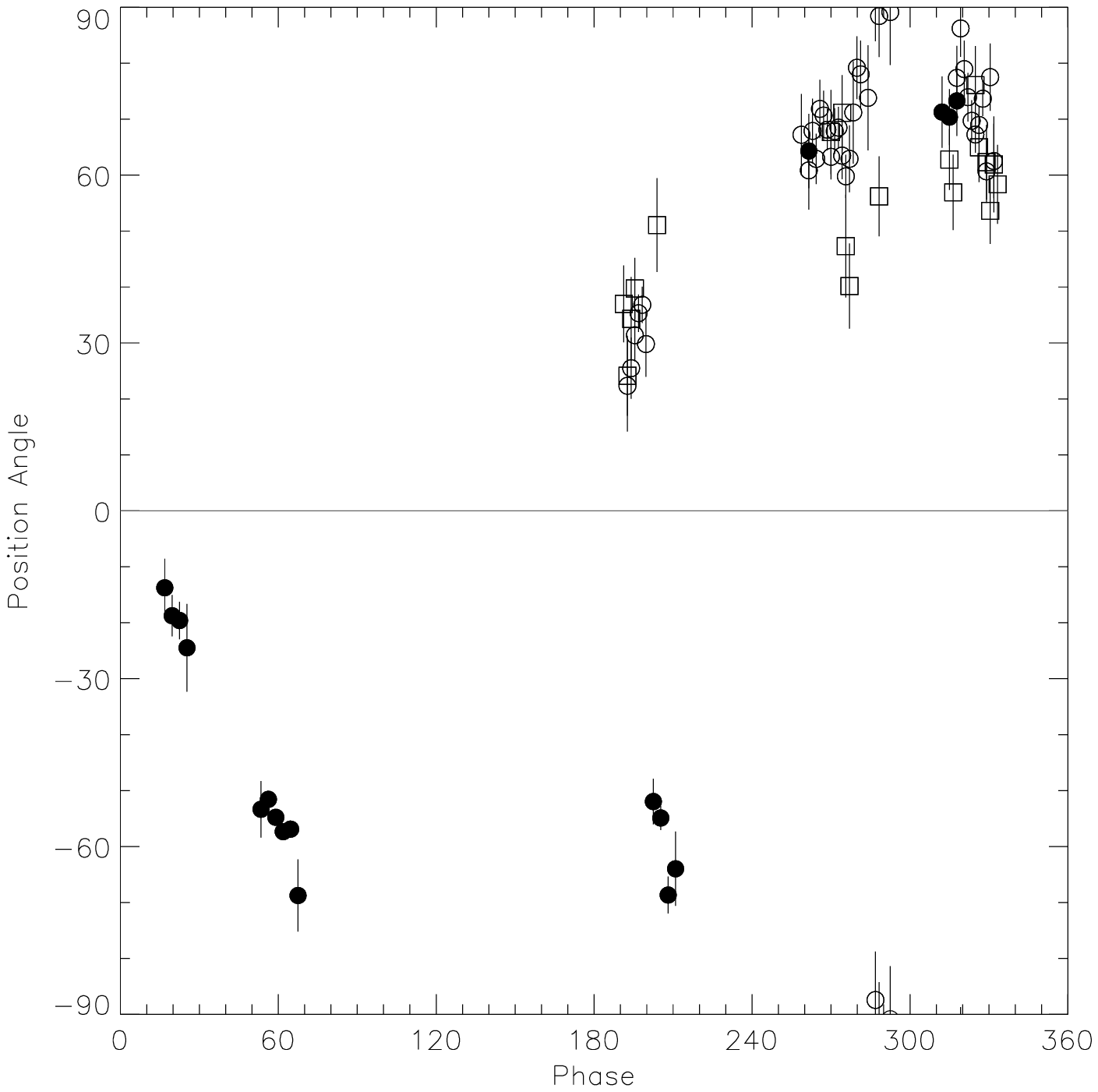}
\figcaption[f4.ps]{Comparison plot of polarization position angles
at 1.424\,GHz (solid circles), 4.885\,GHz (open circles), and
8.435\,GHz (squares).  Note the $90^\circ$ PA separation of the IP
between the 1.4\,GHz and higher frequencies.\label{angle3freq_fig}}

\newpage
\epsscale{1.0}
\plotone{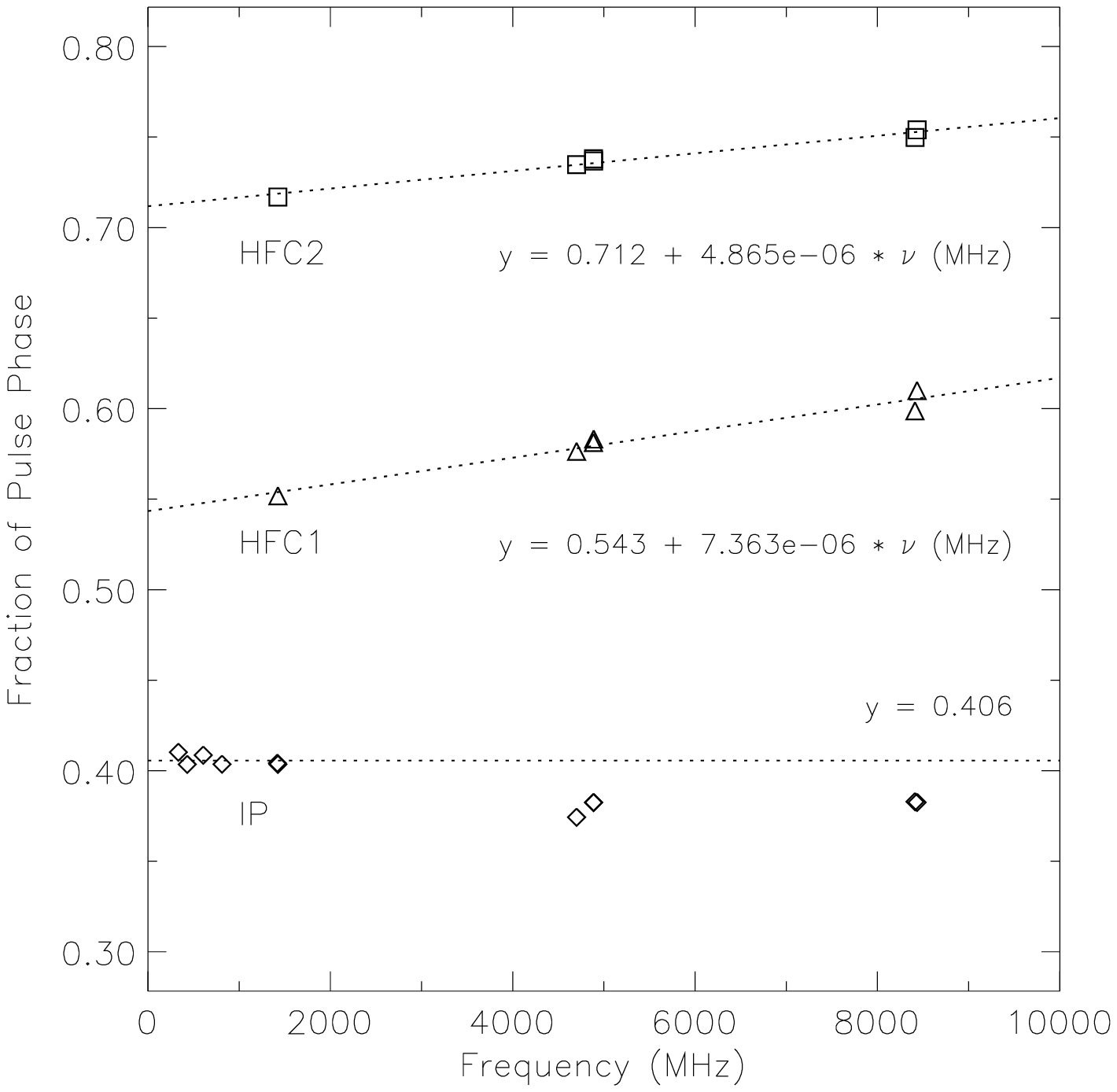}
\figcaption[f5.ps]{Phase separation of selected components from the 
MP with frequency.  MP to IP separation at low frequencies
lie on line drawn at $\delta\phi = 0.406$.  Best fit lines and solutions 
are shown for HFC1 and HFC2.\label{phase_sep_fig}}

\newpage
\epsscale{1.0}
\plotone{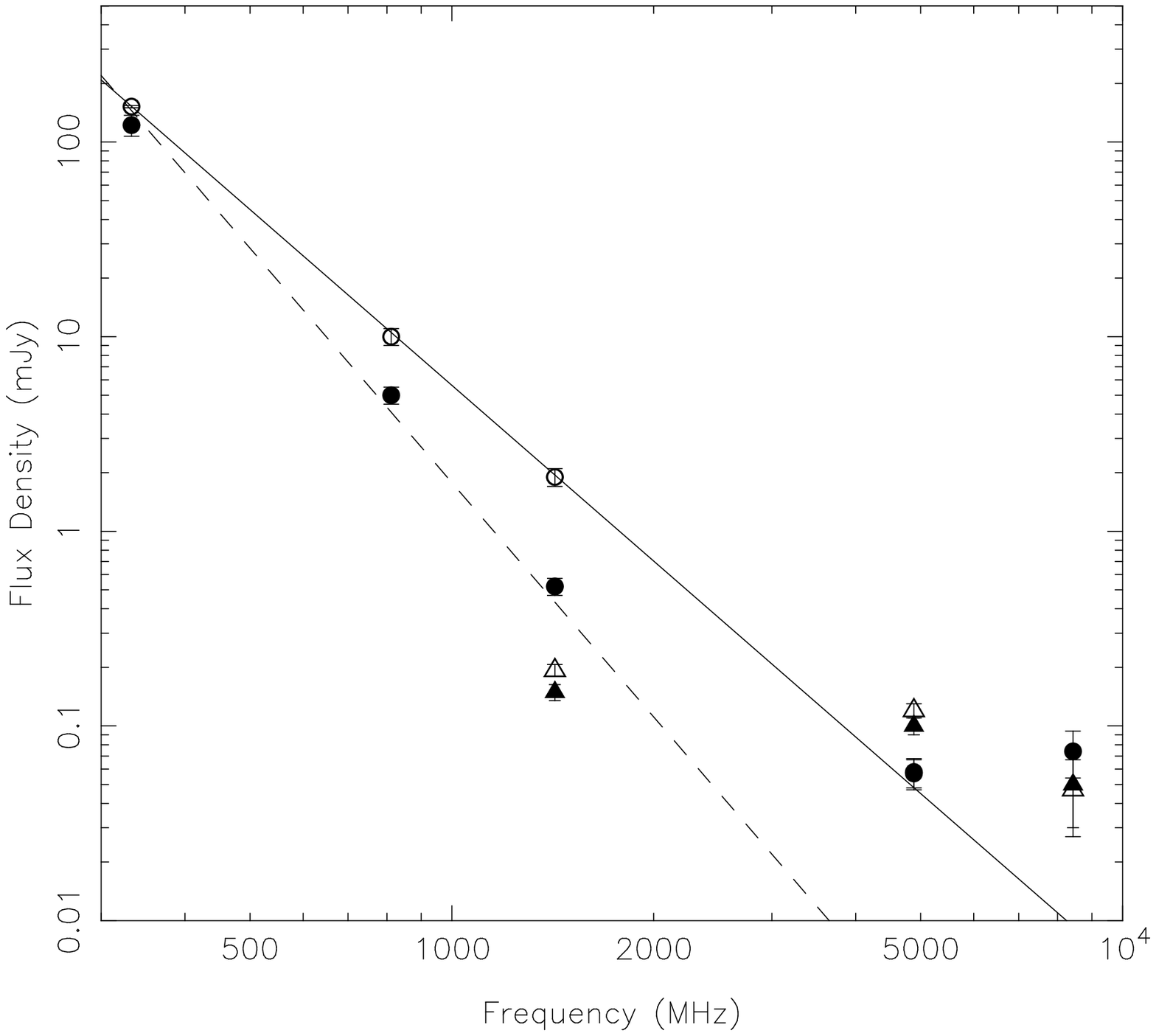}
\figcaption[f6.ps]{Flux density spectrum of the Crab's
major profile components: MP (open circles), IP (filled circles), 
HFC1 (open triangles), and HFC2 (filled triangles).  
The solid line shows the power-law slope of the main pulse's spectral index,
$\alpha = -3$. The dashed line shows the slope of the low-frequency 
interpulse spectral index, $\alpha = -4.1$.\label{flux_vs_freq_fig}}

\newpage
\epsscale{1.0}
\plotone{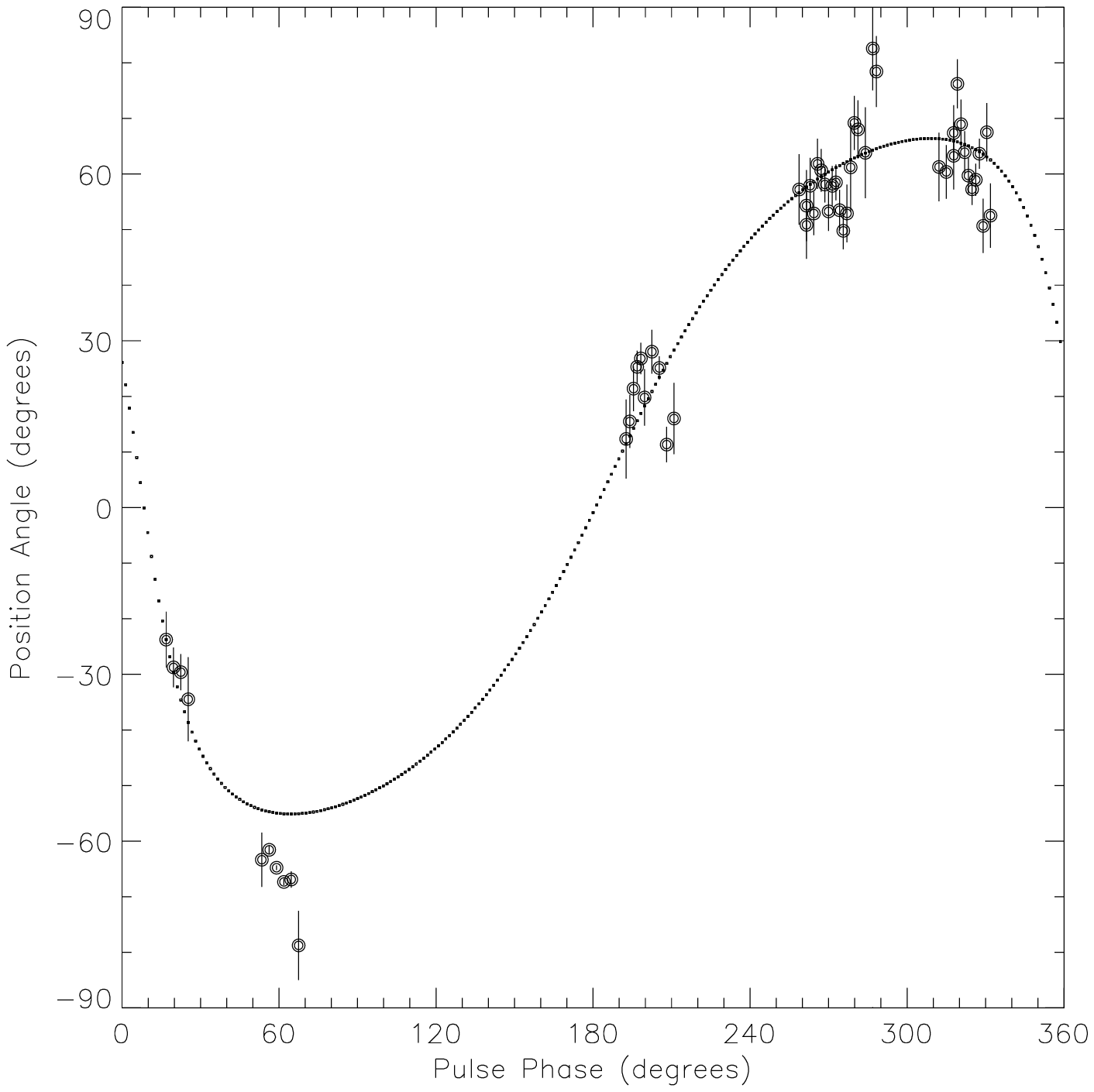}
\figcaption[f7.ps]{Plot of a sample rotating-vector model fit through linear
polarization position angles of the Crab pulsar at 1.4\,GHz (after shifting
the PA of the interpulse by $90^\circ$), 4.9 and 8.4\,GHz.  The RVM fit 
yields an inclination angle, $\alpha = 56.0^\circ$, and a line of sight 
impact angle $\zeta - \alpha = 51.3^\circ$.\label{pafit_fig}}

\end{document}